\documentclass[cameraready]{Interspeech}

\usepackage{newtxtext}
\usepackage{newtxmath}
\usepackage{amsmath,graphicx,hyperref}
\usepackage{booktabs}
\usepackage{multirow}
\usepackage{subcaption}

\newcommand{\etal}{\textit{et al.}}
\newcommand{\y}{\checkmark}
\newcommand{\n}{}

\title{Multi-Loss Learning for Speech Emotion Recognition \\ with Energy-Adaptive Mixup and Frame-Level Attention}

\author[affiliation={1,\dagger,\ddagger}]{Cong}{Wang}
\author[affiliation={1,\dagger,\ddagger}]{Yizhong}{Geng}
\author[affiliation={1}]{Yuhua}{Wen}
\author[affiliation={1}]{Qifei}{Li}
\author[affiliation={1}]{Yingming}{Gao}
\renewcommand{\authorsep}{\\ } 
\author[affiliation={2}]{Ruimin}{Wang}
\renewcommand{\authorsep}{, } 
\author[affiliation={2}]{Chunfeng}{Wang}
\author[affiliation={2}]{Hao}{Li}
\author[affiliation={1,*}]{Ya}{Li}
\author[affiliation={2,*}]{Wei}{Chen}

\address{
    $^{1}$Beijing University of Posts and Telecommunications, Beijing, China \\ 
    $^{2}$Li Auto, Beijing, China
}

\email{yli01@bupt.edu.cn}


\usepackage{comment}

\begin{document}
\maketitle

\begingroup
\renewcommand\thefootnote{}
\footnotetext{$^\dag$ Equal contribution. $^*$ Corresponding authors.}
\footnotetext{$^\ddagger$ Work performed during the internship at Li Auto.}
\endgroup

\begin{abstract}
    Speech emotion recognition (SER) is an important technology in human-computer interaction. However, achieving high performance is challenging due to emotional complexity and scarce annotated data. To tackle these challenges, we propose a multi-loss learning (MLL) framework integrating an energy-adaptive mixup (EAM) method and a frame-level attention module (FLAM). The EAM method leverages SNR-based augmentation to generate diverse speech samples capturing subtle emotional variations. FLAM enhances frame-level feature extraction for multi-frame emotional cues. Our MLL strategy combines Kullback-Leibler divergence, focal, center, and supervised contrastive loss to optimize learning, address class imbalance, and improve feature separability. We evaluate our method on four widely used SER datasets: IEMOCAP, MSP-IMPROV, RAVDESS, and SAVEE. The results demonstrate our method achieves state-of-the-art performance, suggesting its effectiveness and robustness.
\end{abstract}

\section{Introduction}
\label{sec:intro}

Speech emotion recognition (SER) is a important technology in human-computer interaction (HCI), enabling systems to intelligently recognize and respond to human emotions, thereby improving user experiences. SER has broad applications across various domains, including healthcare~\cite{Gumelar2021SR}, customer service~\cite{Prabhakar2023SERCA}, conversational agents~\cite{Hu2023ECA}, and online education~\cite{Vyakaranam2024}. Despite advances, SER remains challenging due to the intrinsically complex and subjective nature of human emotions.

Recent research indicates that emotions in speech are conveyed not only through linguistic content but also through subtle nonverbal cues, such as tone, rhythm, and energy variations~\cite{al2023speech, kaur2023trends}. Accurately modeling these characteristics is essential for capturing robust emotional features and enhancing overall SER performance. However, annotating emotional speech data is a time-consuming and labor-intensive process, inevitably leading to datasets of limited scale. This data scarcity severely restricts the representational learning capacity, thereby hindering the performance of SER systems in real-world scenarios. 

To mitigate data scarcity, researchers have employed data augmentation techniques to improve SER performance. For instance, An et al.~\cite{an2021speech} applied additive noise to expand training corpus, achieving accuracy improvements. More recently, advanced methods such as mixup, have been adopted due to their remarkable effectiveness. Kang et al.~\cite{kang2023learning} introduced a label-adaptive mixup for SER, combining speech to create mixed-label representations. However, this approach ignores energy dynamics in speech by mixing segments uniformly. This simplification that may overlook critical emotional nuances, leading to suboptimal feature representations. 

To address these limitations, we propose a novel SER framework. It integrates an energy-adaptive mixup (EAM) method to generate diverse virtual speech samples with varied energy levels. Concurrently, a frame-level attention module (FLAM) is introduced to refine the extraction of these multi-frame emotional cues. The framework is then jointly optimized by our multi-loss learning (MLL) strategy, which combines four complementary losses: Kullback-Leibler (KL) divergence for soft label alignment, focal loss for hard samples, and both center and supervised contrastive (SupCon) losses to maximize intraclass compactness and interclass separability.

We validate our approach on four widely-used SER datasets: IEMOCAP~\cite{busso2008iemocap}, MSP-IMPROV~\cite{busso2016msp}, RAVDESS~\cite{livingstone2018ryerson}, and SAVEE~\cite{jackson2014surrey}. Comprehensive experimental results demonstrate our method significantly outperforms current SOTA models and exhibits strong generalization across diverse emotional conditions, underscoring its superior robustness. The key contributions of this work are summarized as follows:
\vspace{-0.1cm}
\begin{itemize}
\item We propose a novel EAM method alongside a FLAM pooling method to capture robust emotional representations. To the best of our knowledge, EAM is the first approach to incorporate the energy dynamics of speech signals into the mixup. 
\vspace{-0.45cm}
\item We propose a MLL strategy that, for the first time, integrates the SupCon loss and the center loss for SER. This unified optimization strategy effectively leverages latent emotional features, yielding substantial performance improvements.
\vspace{-0.05cm}
\item Extensive experiments conducted across four datasets validate both the effectiveness and the strong generalization capability of our proposed method. Our approach consistently surpasses existing SOTA models on all datasets, emphasizing not only its superior accuracy but also its remarkable robustness in complex, real-world scenarios.
\end{itemize}
\vspace{-0.1cm}

\begin{figure*}[!ht]
  \centering
  \includegraphics[width=1.0\linewidth]{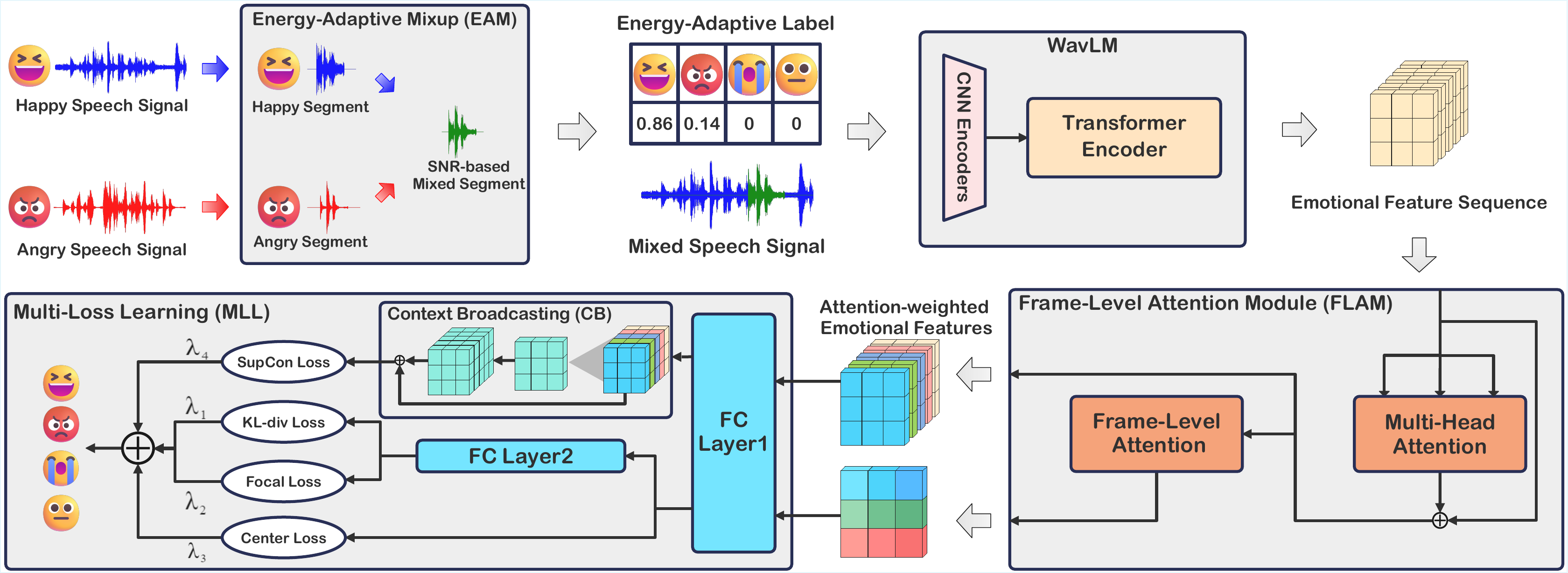}
  \vspace{-0.6cm}
  \caption{The overall model architecture of our proposed SER method.}
  \label{fig:architecture}
\vspace{-0.4cm}
\end{figure*}

\section{Method}
\label{method}
\vspace{-0.1cm}
\subsection{Model architecture}
\vspace{-0.1cm}
As shown in Figure~\ref{fig:architecture}, our model integrates three core components: EAM, FLAM, and MLL strategy. The EAM method enriches the training corpus by generating mixed speech samples with varied energy levels based on signal-to-noise ratio (SNR) adjustments, which provides richer emotional variations. The FLAM then refines temporal relationships between frames using a multi-head attention pooling mechanism to focus on salient emotional cues. Finally, our MLL strategy jointly optimizes the model by combining multiple specialized loss functions (KL-div, focal, center, and SupCon) to handle label distributions, difficult samples, and improve feature separability.

\vspace{-0.3cm}
\subsection{Energy-adaptive mixup method}
\vspace{-0.1cm}
Inspired by the label-adaptive mixup (LAM) method~\cite{kang2023learning}, which ignores energy factors, we propose our EAM method. Unlike the simple length-based label weighting employed in LAM, EAM specifically incorporates the unique energy characteristics of speech signals. 

First, inspired by the overlap algorithm in WavLM~\cite{chen2022wavlm}, we introduce a dynamic segment extraction mechanism. To preserve emotional dominance, the process selects a random mix length $l_{\text{mix}}$ that is constrained to under half the total length $l_i$ of the $i$-th original sample. Furthermore, to significantly enhance data diversity, starting positions $s_i$ and $s_j$ are randomly selected to extract speech segments $\mathbf{x}'_i=\mathbf{x}_i[s_i:s_i + l_{\text{mix}}]$ and $\mathbf{x}'_j=\mathbf{x}_j[s_j:s_j + l_{\text{mix}}]$ for the subsequent mixing operation. 

Second, to simulate complex emotional interference in real-world scenarios and expand the energy distribution of the training data, we design an SNR-based energy adjustment mechanism rather than directly mixing the raw segments. Treating the interfering segment $\mathbf{x}'_j$ as noise, its energy $P'_j$ is scaled to match a randomly sampled SNR value (in dB). Specifically, the scaling factor $\textit{scale}$ and the energy-adjusted segment $\mathbf{x}''_j$, with its new energy $P''_j$, are computed as:
\begin{align}
    \textit{scale} &= \sqrt{\frac{P'_i}{10^{\text{SNR}/10} \cdot P'_j}}, \tag{1} \\
    \mathbf{x}''_j &= \textit{scale} \cdot \mathbf{x}'_j. \tag{2}
\end{align}
where $P'_i$ denotes the energy of $\mathbf{x}'_i$, and $\text{SNR} \in [-5,10]$. 

Finally, the adjusted segment $\mathbf{x}''_j$ is overlaid onto the dominant segment $\mathbf{x}'_i$:
\begin{equation}
    \mathbf{x}_{\text{mix}}[s_i : s_i + l_{\text{mix}}] = \mathbf{x}'_i + \mathbf{x}''_j. \tag{3}
\end{equation}
To accurately reflect this acoustic mixture, our proposed energy-adaptive method dynamically calculates a weight $\lambda_{\text{mix}}$ using instantaneous energy and temporal coverage ratios to compute the final soft label $\mathbf{y}_{\text{mix}}$:
\begin{equation}
  \lambda_{\text{mix}} = \frac{P''_j}{P'_i + P''_j} \cdot \frac{l_{\text{mix}}}{l_i}, \quad \mathbf{y}_{\text{mix}} = (1-\lambda_{\text{mix}})\mathbf{y}_i + \lambda_{\text{mix}}\mathbf{y}_j. \tag{4}
\end{equation}
Subsequently, the mixed signal $\mathbf{x}_{\text{mix}}$ is fed into a pre-trained WavLM~\cite{chen2022wavlm} to extract the emotional feature sequence $\mathbf{F} \in \mathbb{R}^{T \times D}$, where $T$ denotes the number of frames and $D$ is the feature dimension. By generating richer samples and a more representative label distribution, EAM better captures the correlation between energy and emotion~\cite{el2011survey}. 

\vspace{-0.3cm}
\subsection{Frame-level attention module}
\vspace{-0.1cm}

Our FLAM captures subtle temporal dependencies by enhancing inter-frame relationships. The input emotional feature sequence $\mathbf{F} \in \mathbb{R}^{T \times D}$ is first processed by a 16-head Multi-Head Self-Attention (MSA) module with a residual connection, yielding $\mathbf{F}' = \mathbf{F} + \text{MSA}(\mathbf{F})$. To aggregate the frame-level sequence $\mathbf{F}'$ into a single utterance-level feature vector $\mathbf{f} \in \mathbb{R}^{D}$, we propose an attention pooling mechanism. Unlike traditional mean or max pooling, which uniformly dilutes salient emotional cues or indiscriminately discards contextual details, our approach dynamically weighs frame importance. By introducing a learnable projection vector $\mathbf{w} \in \mathbb{R}^{D \times 1}$, the aggregated robust feature $\mathbf{f}$ is computed as:
\begin{equation}
     \alpha_t = \frac{\exp({\mathbf{f}'_t}^\top \mathbf{w})}{\sum_{i=1}^T \exp({\mathbf{f}'_i}^\top \mathbf{w})}, \quad \mathbf{f} = \sum_{t=1}^{T} \alpha_t \mathbf{f}'_t. \tag{5}
\end{equation}
where $\mathbf{f}'_t \in \mathbb{R}^{D}$ represents the feature vector of the $t$-th frame in $\mathbf{F}'$, and $\alpha_t$ is the normalized attention weight for that frame. This targeted aggregation allows the model to actively focus on the most emotionally discriminative frames, yielding superior representations as empirically demonstrated in our evaluations.

\vspace{-0.3cm}
\subsection{Multi-loss learning strategy}
\vspace{-0.1cm}

We employ a weighted MLL strategy to optimize the model. Let $C$ denote the total number of emotion classes. Since the EAM label $\mathbf{y}_{\text{mix}}$ serves as a soft target distribution, we define the probability of the $c$-th class as $y_c$ and the model's predicted probability as $\hat{y}_c$. To measure the distribution discrepancy and focus on hard-to-classify samples, we combine KL-divergence and Focal loss~\cite{lin2017focal}:
\begin{align}
\mathcal{L}_{\text{KL}} &= \sum_{c=1}^{C} y_c \log\left(\frac{y_c}{\hat{y}_c}\right), \tag{6} \\
\mathcal{L}_{\text{Focal}} &= - \sum_{c=1}^{C} y_c (1 - \hat{y}_c)^\gamma \log(\hat{y}_c), \tag{7}
\end{align}
where $\gamma$ is the focusing parameter. 

To improve feature discrimination, we use a linear layer to project the utterance-level aggregated feature $\mathbf{f}$ into a lower-dimensional representation $\mathbf{z}$. The center loss~\cite{wen2016discriminative} minimizes intra-class variance within a batch of size $B$:
\begin{equation}
\mathcal{L}_{\text{Center}} = \frac{1}{B} \sum_{b=1}^{B} \| \mathbf{z}_b - \mathbf{c}_{{y}_b} \|^2, \tag{8}
\end{equation}
where $\mathbf{z}_b$ is the projected feature of the $b$-th sample and $\mathbf{c}_{y_b}$ is its corresponding class center. 

Furthermore, for frame-level features, we project the sequence $\mathbf{F}'$ into a lower dimension. Let $\mathbf{h}_t$ be the projected feature at frame $t \in \{1, \dots, T\}$. We adapt a context broadcasting (CB) mechanism~\cite{hyeon2023scratching} to encourage sparse feature interactions, yielding the enhanced frame feature $\tilde{\mathbf{h}}_t$:
\begin{equation}
\tilde{\mathbf{h}}_t = \frac{1}{2}\left(\mathbf{h}_t + \frac{1}{T} \sum_{m=1}^{T} \mathbf{h}_m\right). \tag{9}
\end{equation}  
Finally, we apply supervised contrastive (SupCon) loss~\cite{khosla2020supervised} over the flattened set $I$ of all $B \times T$ frames in the batch to maximize inter-class distance:
\begin{equation}
    \mathcal{L}_{\text{SupCon}} = \frac{1}{B \times T} \sum\limits_{i \in I} \frac{-1}{|P(i)|} \sum\limits_{j \in P(i)} \log \frac{ \exp(\tilde{\mathbf{h}}_i^\top \tilde{\mathbf{h}}_j / \tau)}{\sum\limits_{k \in A(i)} \exp(\tilde{\mathbf{h}}_i^\top \tilde{\mathbf{h}}_k / \tau)}, \tag{10}
\end{equation}
where $\tau$ is the temperature scalar, $P(i)$ is the set of positive samples for $i$, and $A(i)$ contains all samples in $I$ excluding $i$. 

The final objective is a weighted summation of these components:
\begin{equation}
\mathcal{L} = \lambda_1 \mathcal{L}_\text{KL} + \lambda_2 \mathcal{L}_\text{Focal} + \lambda_3 \mathcal{L}_\text{Center} + \lambda_4 \mathcal{L}_\text{SupCon}. \tag{11}
\end{equation}
where $\lambda_1, \dots, \lambda_4$ are scaling factors empirically set to normalize each weighted loss component into the $[0, 1]$ range, thereby preventing gradients from being dominated by any single objective and ensuring a balanced optimization process.

\section{Experiments and results}

\label{experiments}

\subsection{Datasets}
We evaluate on four SER datasets: IEMOCAP~\cite{busso2008iemocap}, MSP-IMPROV~\cite{busso2016msp}, RAVDESS~\cite{livingstone2018ryerson}, and SAVEE~\cite{jackson2014surrey}. Covering spontaneous and acted emotions across diverse conditions, they provide a robust testbed for speaker-independent SER. 

\textbf{IEMOCAP}~\cite{busso2008iemocap} contains 12 hours of spontaneous and acted conversational speech from 10 speakers in five sessions. Merging ``excited'' with ``happy'' yields 5,531 utterances across four emotions (happy: 1,636, angry: 1,103, sad: 1,084, neutral: 1,708), evaluated via 5-fold session-independent Cross-Validation (CV). 

\textbf{MSP-IMPROV}~\cite{busso2016msp} contains 8,438 spontaneous clips from 12 actors in six sessions. Using the same four basic emotions, we conduct 6-fold session-independent CV. 

\textbf{RAVDESS}~\cite{livingstone2018ryerson} comprises 1,440 acted speech utterances from 24 actors covering eight emotions (happy, angry, sad, neutral, fearful, disgust, surprised, calm). We perform 6-fold subject-independent CV. 

\textbf{SAVEE}~\cite{jackson2014surrey} contains 480 utterances from four male speakers across seven emotions (excluding calm). We perform 4-fold speaker-independent CV and report speaker-wise UA to demonstrate robustness against speaker variability. 

\begin{table}[ht!]
    \caption{Comparison results on IEMOCAP, MSP-IMPROV, and RAVDESS datasets (A: Audio-only, M: Multi-modal).}
    \vspace{-0.35cm}
    \label{tab:sota-vertical}
    \centering
    \setlength{\tabcolsep}{2pt} 
    \scalebox{0.9}{
    \begin{tabular}{llccc}
    \toprule
    \textbf{Dataset} & \textbf{Methods} & \textbf{Modality} & \textbf{WA(\%)} & \textbf{UA(\%)} \\
    \midrule
    \multirow{8}{*}{\textbf{IEMOCAP}} 
    & Tang~\etal~\cite{tang2025speech} & A & 71.64 & 72.72 \\
    & Sun~\etal~\cite{sun2024combining} & A & 72.86 & 72.85 \\
    & Wang~\etal~\cite{wang2025speech} & A & 73.37 & 74.18 \\
    & He~\etal~\cite{he2023multiple} & A & 73.80 & 74.25 \\
    & Gao~\etal~\cite{gao2023two} & A & 74.94 & 76.10 \\
    & Kang~\etal~\cite{kang2023learning} & A & 75.37 & 76.04 \\
    \cmidrule{2-5}
    & He~\etal~\cite{he2023multilevel} & M & 74.50 & 75.00 \\
    & Wang~\etal~\cite{wang2023exploring} & M & 75.20 & 76.40 \\
    \cmidrule{2-5}
    & \textbf{Ours} & A & \textbf{78.47} & \textbf{79.14} \\
    \midrule
    \multirow{7}{*}{\textbf{MSP-IMPROV}} 
    & Guo~\etal~\cite{guo2021representation} & A & 46.20 & 44.70 \\
    & Nediyanchath~\etal~\cite{nediyanchath2020multi} & A & 47.30 & 46.10 \\
    & Xu~\etal~\cite{xu2021speech} & A & 47.90 & 45.80 \\
    & Cao~\etal~\cite{cao2021hierarchical} & A & 50.70 & 49.90 \\
    & Liu~\etal~\cite{liu2024contrastive} & A & 51.51 & 41.56 \\
    & Liu~\etal~\cite{liu2022atda} & A & 55.80 & 55.30 \\
    \cmidrule{2-5}
    & \textbf{Ours} & A & \textbf{58.55} & \textbf{58.34} \\
    \midrule
    \multirow{8}{*}{\textbf{RAVDESS}} 
    & Baevski~\etal~\cite{baevski2020wav2vec} & A & 74.38 & 73.44 \\
    
    & Sun~\etal~\cite{sun2024hicmae} & A & 72.29 & 70.38 \\
    & Chen~\etal~\cite{chen2022wavlm} & A & 75.36 & 75.28 \\
    & Yu~\etal~\cite{yu2024speech} & A & 81.86 & 82.75 \\
    \cmidrule{2-5}
    & Chumachenko~\etal~\cite{chumachenko2022self} & M & 79.20 & - \\
    & Sadok~\etal~\cite{sadok2023vector} & M & 84.80 & - \\
    & Sun~\etal~\cite{sun2024hicmae} & M & 87.99 & 87.96 \\
    \cmidrule{2-5}
    & \textbf{Ours} & A & \textbf{93.40} & \textbf{92.28} \\
    \bottomrule
    \end{tabular}
    }
\vspace{-0.6cm}
\end{table}

\begin{table*}[ht!]
    \caption{Comparison results on SAVEE in UA(\%).}
    \vspace{-0.35cm}
    \label{tab:sota-SAVEE}
    \centering
    \scalebox{1.0}{
    \begin{tabular*}{0.9\textwidth}{@{\extracolsep{\fill}}llllll}
    \toprule
    \textbf{Methods} & \textbf{DC} & \textbf{JE} & \textbf{JK} & \textbf{KL} & \textbf{Mean}\\
    \midrule
    Chen~\etal~\cite{chen2023exploring} & 81.6 & 83.3 & \textbf{69.9} & 49.7 & 71.1\\
    \textbf{Ours} & \textbf{82.3} & \textbf{87.0} & 66.9 & \textbf{53.0} & \textbf{72.3}\\
    \midrule
    Human~\cite{chen2023exploring} & 73.7 & 67.7 & 71.2 & 53.2 & 66.5\\
    \bottomrule
    \end{tabular*}
    }
\end{table*}

\subsection{Experiment setup}
We evaluate our method in a speaker-independent setting using unweighted accuracy (UA) and weighted accuracy (WA). Ablation studies isolate the contributions of key components (EAM, FLAM) and individual losses (Focal, SupCon) under identical configurations. For implementation, Center and SupCon losses operate in a 64-dimensional space. Models are trained on an NVIDIA RTX 3090 GPU with a batch size of 16 using the Adam optimizer. The initial learning rates are $1 \times 10^{-4}$ for the model and $5 \times 10^{-3}$ for center updates, both decaying by a factor of 7/8 per epoch until the 20th epoch.

\begin{table*}[t!h]
    \caption{Ablation study of the proposed method on IEMOCAP, including pre-trained model, mixup method, feature aggregation method, and loss function. We use Kang et al.'s method~\cite{kang2023learning} as the baseline model, as shown in the first row of the table.}
    \vspace{-0.35cm}
    \label{tab:ablation}
    \centering
    \scalebox{0.9}{
    \begin{tabular}{cc|cc|ccc|cccc|cc}
    \toprule
    Hubert & \textbf{WavLM} &
    \multicolumn{2}{c|}{\textbf{Mixup}} &
    \multicolumn{3}{c|}{\textbf{Feature Aggregation}} &
    \multicolumn{4}{c|}{\textbf{Loss Function}} & 
    \multirow{2}{*}{\textbf{WA(\%)}} & 
    \multirow{2}{*}{\textbf{UA(\%)}} \\ 
    Large&\textbf{Large} & LAM&\textbf{EAM} &
    MaxPool&MeanPool&\textbf{FLAM}& KL-div&Center&Focal&SupCon & & \\ 
    \hline
    \y&\n & \y&\n & \n&\n&\n & \y&\y&\n&\n & 75.37 & 76.04 \\
    \y&\n & \n&\n & \n&\n&\y & \y&\y&\n&\n & 75.83 & 76.45 \\
    \y&\n & \n&\y & \n&\n&\n & \y&\y&\n&\n & 76.18 & 76.69 \\
    \hline
    \y&\n & \y&\n & \n&\n&\y & \y&\y&\n&\n & 76.22 & 76.84 \\
    \y&\n & \n&\y & \n&\n&\y & \y&\y&\n&\n & 76.31 & 76.90 \\
    \hline
    \n&\y & \y&\n & \n&\n&\n & \y&\y&\n&\n & 76.98 & 77.14 \\
    \n&\y & \n&\n & \n&\n&\y & \y&\y&\n&\n & 77.12 & 77.58 \\
    \n&\y & \n&\y & \n&\n&\n & \y&\y&\n&\n & 77.26 & 77.71 \\
    \hline
    \n&\y & \n&\y & \y&\n&\n & \y&\y&\n&\n & 77.32 & 77.74 \\
    \n&\y & \n&\y & \n&\y&\n & \y&\y&\n&\n & 77.47 & 77.95 \\
    \n&\y & \y&\n & \n&\n&\y & \y&\y&\n&\n & 77.58 & 78.02 \\
    \n&\y & \n&\y & \n&\n&\y & \y&\y&\n&\n & 77.63 & 78.10 \\
    \hline
    \n&\y & \n&\y & \n&\n&\y & \y&\y&\y&\n & 77.96 & 78.56 \\
    \n&\y & \n&\y & \n&\n&\y & \y&\y&\n&\y & 78.23 & 78.83 \\
    \n&\y & \n&\y & \n&\n&\y & \y&\y&\y&\y & \textbf{78.47} & \textbf{79.14} \\
    \bottomrule
    \end{tabular}
    }
\end{table*} 

\subsection{Results and discussions}
Our proposed method consistently outperforms existing SOTA approaches across all benchmarks, exhibiting robust generalization under both spontaneous and acted emotional conditions. 

On \textbf{IEMOCAP} (Table~\ref{tab:sota-vertical}), our method achieves 78.47\% WA and 79.14\% UA. It not only yields clear gains over the strong audio-only LAM baseline (Kang~\etal~\cite{kang2023learning}) but also surpasses recent multi-modal approaches. For \textbf{MSP-IMPROV}, we obtain 58.55\% WA and 58.34\% UA, outperforming the top baseline (Liu~\etal~\cite{liu2022atda}) by a notable 3.04\% UA. These results demonstrate that unlike traditional mixup, our energy-aware EAM effectively models the nuanced acoustic variations in spontaneous speech. 

For acted datasets, the performance improvements are exceptionally significant. On \textbf{RAVDESS}, we reach 93.40\% WA and 92.28\% UA, outperforming top audio-only baselines and even several multi-modal approaches (Table~\ref{tab:sota-vertical}). This validates that explicitly modeling energy distributions aligns perfectly with the pronounced prosodic and intensity patterns typical of acted emotional speech. Furthermore, on \textbf{SAVEE} (Table~\ref{tab:sota-SAVEE}), we achieve an average UA of 72.3\%, observing consistent improvements across challenging speakers. This demonstrates that the proposed training strategy generalizes robustly under strong speaker variability. 

Overall, these substantial gains stem from the synergy of our three core innovations: EAM generating physically grounded, energy-diverse samples to enrich the feature space; FLAM dynamically aggregating emotionally salient frames rather than indiscriminately pooling them; and MLL optimizing feature compactness and separability. 

\begin{figure}[t!]
    \centering
    \begin{subfigure}[b]{0.48\columnwidth}
        \centering
        (a)
        \includegraphics[width=\linewidth]{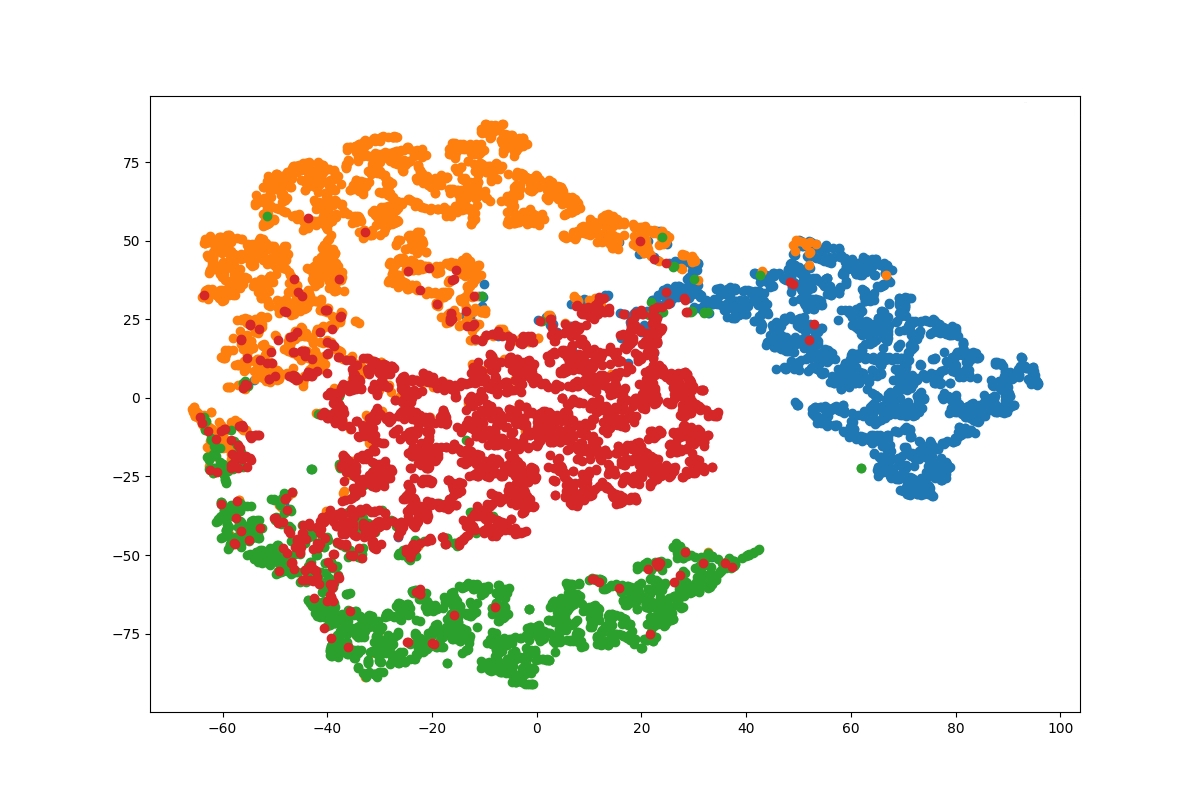}
    \end{subfigure}
    \begin{subfigure}[b]{0.48\columnwidth}
        \centering
        (b)
        \includegraphics[width=\linewidth]{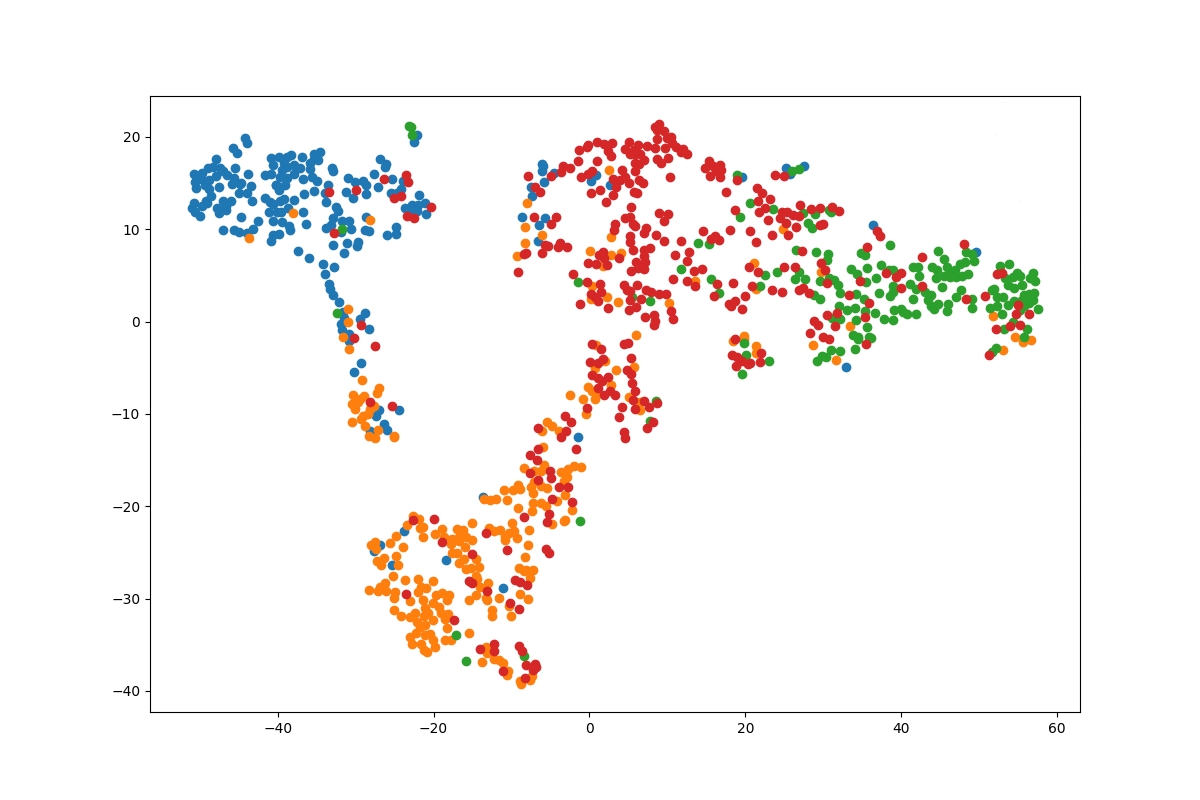}
    \end{subfigure}
    \begin{subfigure}[b]{0.48\columnwidth}
        \centering
        (c)
        \includegraphics[width=\linewidth]{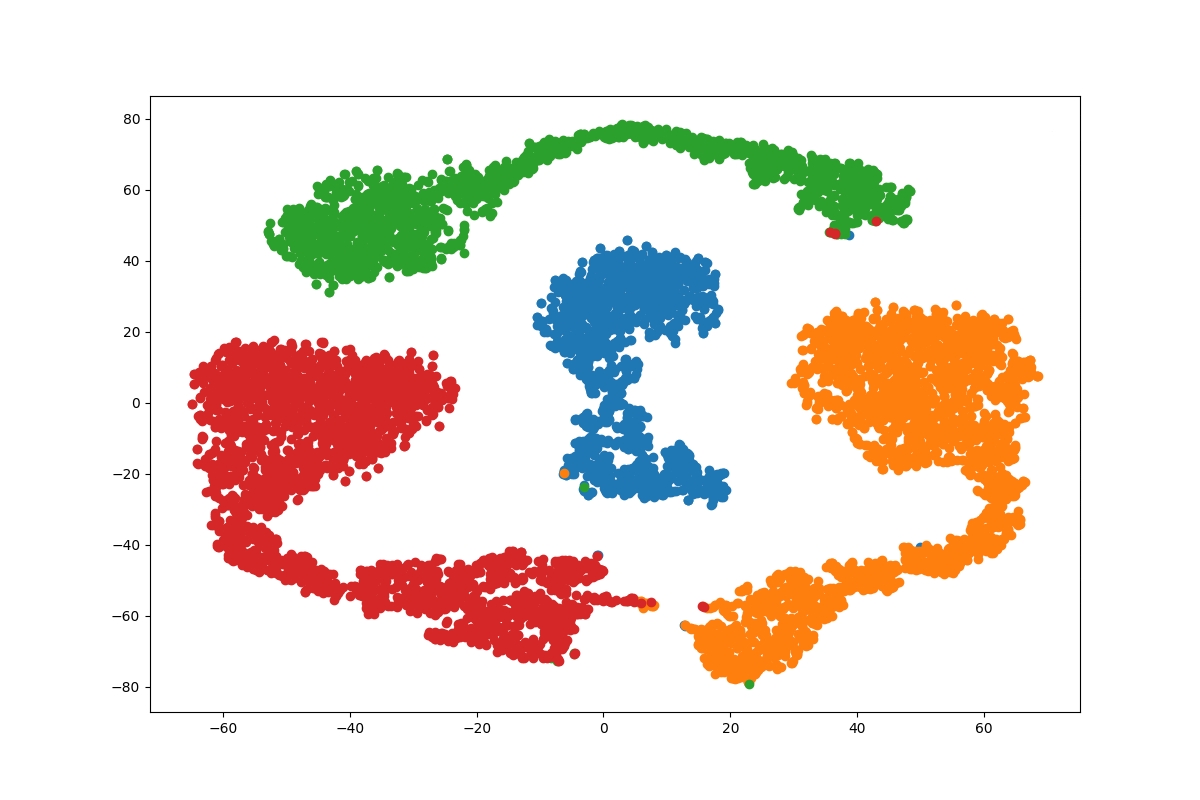}
    \end{subfigure}
    \begin{subfigure}[b]{0.48\columnwidth}
        \centering
        (d)
        \includegraphics[width=\linewidth]{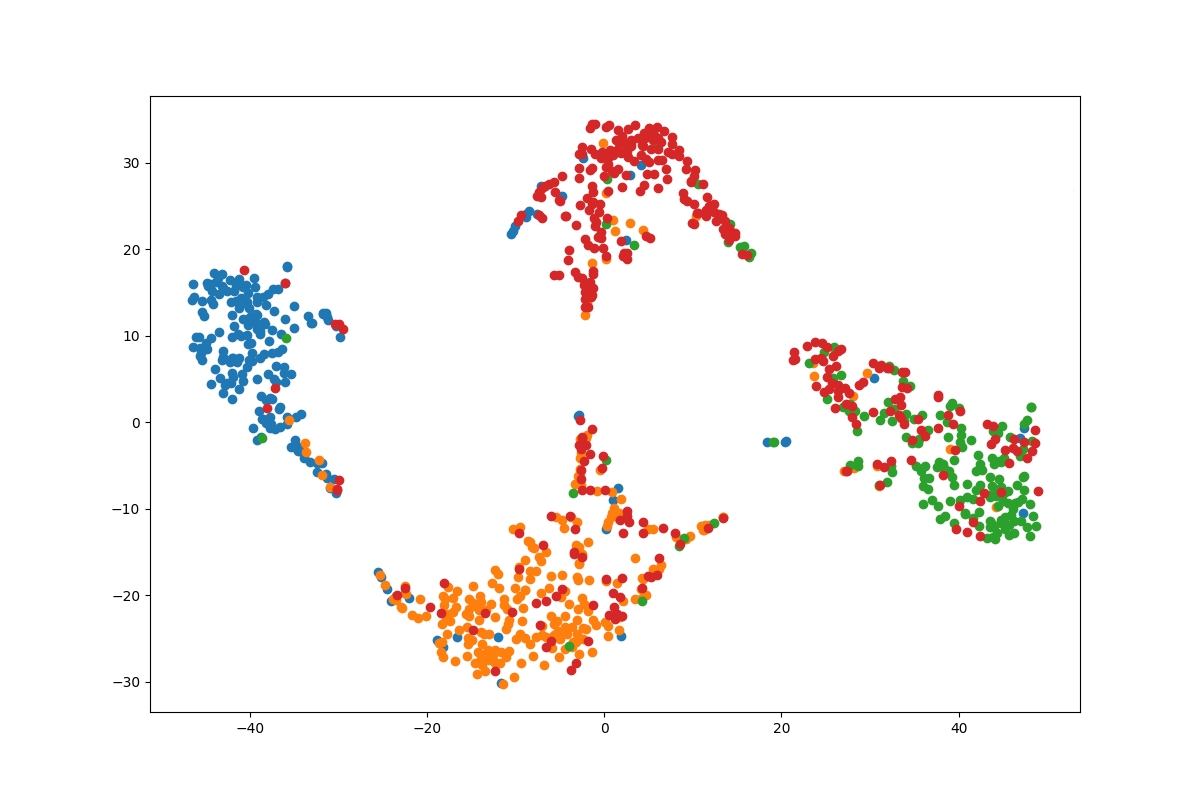}
    \end{subfigure}
    \vspace{-0.25cm}
    \caption{t-SNE visualizations of feature distributions on IEMOCAP. (a) Training set before MLL; (b) Test set before MLL; (c) Training set after MLL; (d) Test set after MLL. Colors: Blue–Angry, Orange–Happy, Green–Sad, Red–Neutral. The feature clusters are visibly more distinct after our MLL strategy.}
    \label{TSNE}
\end{figure}

\subsection{Ablation study}
To validate the contribution of each component, we conduct a systematic ablation study on the IEMOCAP dataset (Table~\ref{tab:ablation}). 

First, we confirm the individual effectiveness of EAM and FLAM. Substituting the baseline LAM with our EAM yields consistent improvements, validating that generating energy-diverse samples provides richer representations than mere length-based mixing. Furthermore, FLAM dynamically weighs multi-frame emotional features, achieving 78.10\% UA. This significantly outperforms traditional aggregation methods like MaxPool (77.74\% UA) and MeanPool (77.95\% UA), proving the necessity of targeted frame-level attention. 

Next, we analyze the components of our MLL strategy. Building upon the base KL and center losses, adding focal loss to emphasize hard-to-classify samples improves the UA to 78.56\%. Alternatively, applying SupCon loss to maximize inter-class distance and minimize intra-class variance reaches 78.83\% UA. Combining all four objectives achieves the peak performance of 79.14\% UA. Finally, t-SNE visualizations (Figure~\ref{TSNE}) directly corroborate the MLL strategy's effectiveness, demonstrating noticeably more compact and separable feature clusters compared to the pre-MLL state. 

\section{Conclusion}
\label{conclusion}

In this paper, we propose a novel multi-loss learning (MLL) framework for robust speech emotion recognition (SER). We introduce an energy-adaptive mixup (EAM) to generate energy-diverse samples via SNR-guided adjustment, alongside a frame-level attention module (FLAM) to dynamically aggregate salient temporal features. The framework is optimized by jointly applying KL-divergence, focal, center, and supervised contrastive losses, effectively aligning soft labels, mitigating class imbalance, and enhancing feature discriminability. Experiments on four benchmark datasets demonstrate SOTA performance, with ablations confirming each component's efficacy. Overall, our framework provides a reliable SER solution under limited data and diverse conditions. Future work will extend it to cross-lingual settings, multimodal cues, and more advanced adaptive augmentations.

\section{Generative AI Use Disclosure}

During the preparation of this manuscript, the authors used generative AI tools (e.g., ChatGPT) for language refinement and minor editing. All outputs were carefully reviewed and revised by the authors, who take full responsibility for the content of this paper.

\bibliographystyle{IEEEtran}
\bibliography{mybib}

\end{document}